\documentclass[conference]{IEEEtran}
\IEEEoverridecommandlockouts
\usepackage{cite}
\usepackage[table]{xcolor}
\usepackage{tikz}
\usetikzlibrary{arrows.meta,shapes.geometric}
\usepackage{amsmath,amssymb,amsfonts}
\usepackage{algorithmic}
\usepackage{graphicx}
\usepackage{makecell}
\usepackage{textcomp}

\usepackage{booktabs}

\usepackage{forest}
\def\BibTeX{{\rm B\kern-.05em{\sc i\kern-.025em b}\kern-.08em
    T\kern-.1667em\lower.7ex\hbox{E}\kern-.125emX}}

\usetikzlibrary{shapes.geometric, arrows.meta, positioning, calc}

\definecolor{myblue}{HTML}{2F5597}
\definecolor{myorange}{HTML}{E97132}
\renewcommand{\baselinestretch}{0.95}
\allowdisplaybreaks 
\definecolor{blue}{rgb}{0,0,0}
\definecolor{purple}{rgb}{0,0,0}
\begin{document}

\title{Voltage-Feedback Backoff and Restoration for Priority-Based EV Charging Control}
\author{
\IEEEauthorblockN{Eyad Alsulayhim, Ning Lu}
\IEEEauthorblockA{
\textit{North Carolina State University} \\
\textit{Raleigh, NC 27606, USA} \\
\{efalsula, nlu2\}@ncsu.edu
}
}

\maketitle

\color{purple}
\begin{abstract}
Uncoordinated residential electric vehicle (EV) charging can cause sustained undervoltage in distribution networks. This paper proposes a centralized voltage-feedback backoff and restoration (VFBR) framework that separates concurrent charging-count control from EV selection. The controller adjusts the allowable charging count using monitored critical-bus voltages, without load forecasts, online power-flow calculations, or optimization. Initial charging suspension relieves undervoltage, followed by group-wise restoration and backoff as voltage conditions permit. A complementary priority score combines charging urgency with voltage sensitivity precomputed offline to balance time-critical requests against their location-dependent voltage impacts. One-week simulations on the unbalanced IEEE 123-bus feeder with 500 EVs show that VFBR reduces the proportion of charging-period bus-time samples below 0.95~p.u. from 15\% to less than 1\%. Among the evaluated strategies, combined allocation with urgency and sensitivity weights of 0.75 and 0.25 delivers 25\% more energy than random allocation and achieves the highest mean departure state of charge and fairness. These results demonstrate that simple voltage-feedback control, coupled with urgency- and sensitivity-based allocation, effectively mitigates undervoltage while improving charging service.
\end{abstract}


\section{Introduction}

Coincident residential electric vehicle (EV) charging can increase distribution feeder loading and cause sustained undervoltage, particularly during household demand peaks~\cite{b2,pohl2025}. Although voltage regulators and capacitor banks provide voltage support, coordinated charging directly reduces the demand contributing to voltage drops. A practical charging controller should respond to changing feeder conditions, restore charging when conditions permit, and balance individual charging requirements with network constraints.

Optimization-based approaches explicitly incorporate network constraints and user requirements into charging schedules. For example, Ba\~{n}ol Arias et al.\ proposed a hierarchical mixed-integer linear programming framework considering user satisfaction and unexpected EV arrivals~\cite{banol2023}. Such methods provide systematic allocation but require optimization models and solver execution as schedules are updated. For corrective undervoltage management, measurement-based feedback offers an alternative that adjusts charging according to observed conditions without repeatedly solving a scheduling problem.

Existing feedback methods demonstrate the feasibility of low-complexity charging control. Ucer et al.\ developed a decentralized additive-increase multiplicative-decrease method~\cite{ucer2021}, applying congestion-control principles to EV charging. Quir\'{o}s-Tort\'{o}s et al.\ proposed centralized disconnection and reconnection of charging points for thermal and voltage management, using accumulated charging time to rank EVs~\cite{quiros2016}. These studies establish that charging can be curtailed and restored using limited information. However, accumulated charging time does not explicitly capture the remaining energy requirement relative to the time available before departure.

Charging allocation must also account for the electrical impact of individual requests. Dreucci et al.\ developed centralized congestion-management schemes using power transfer distribution factors, considering electrical impact, fairness, and charging priority~\cite{dreucci2025}. For undervoltage management, the corresponding allocation decision depends on how charging at each location affects critical-bus voltages. Urgency-only allocation may prioritize EVs that consume substantial voltage margin, limiting opportunities for other vehicles. Sensitivity-only allocation favors electrically advantageous locations but may defer time-critical requests. This motivates combining charging urgency and voltage sensitivity within a feedback-based controller.

This paper proposes a voltage-feedback backoff and restoration (VFBR) framework that separates how many EVs may charge from which EVs are selected. Its primary contribution is a controller that adapts the concurrent charging count directly from monitored critical-bus voltages, without load forecasts, forecasts of future EV arrivals, or online network-model calculations and optimization. Initial charging suspension relieves undervoltage, while group-wise restoration and backoff progressively recover charging service as voltage conditions permit. A complementary priority scheme combines charging urgency with voltage sensitivities precomputed offline from the feeder model, balancing time-critical requests against their impact on the feeder's ability to accommodate other EVs. 

One-week simulations on the unbalanced IEEE 123-bus feeder evaluate the resulting tradeoffs among voltage mitigation, charging service, and fairness. The results show that combined allocation improves aggregate charging service, whereas urgency-only allocation better protects the minimum departure SOC. 

\color{black}
\section{Methodology}
\label{sec:methodology}

\subsection{Control Assumptions and Framework Overview}

We assume that a central controller can directly enable or suspend charging for each participating EV connected to the distribution feeder. At discrete intervals of $\Delta t=15$~min, the controller collects voltage measurements from monitored critical buses and updates the charging requests, SOCs, and departure times of connected EVs. Using this information, it determines the allowable number of concurrently charging EVs, $N_{\mathrm{allow}}(t)$, and updates individual charging commands according to a priority ranking. Each selected EV charges at its rated power, while charging for unselected EVs is suspended until a subsequent update.

The controller uses the minimum measured voltage among the monitored critical buses, denoted by $V_{\min}(t)$, to detect undervoltage. A violation occurs when $V_{\min}(t)<V_{\mathrm{lim}}$, where $V_{\mathrm{lim}}=0.95$~p.u. At each interval, the number of EVs commanded to charge cannot exceed $N_{\mathrm{allow}}(t)$ or the number of current charging requests. Voltage feedback determines the allowable number, while charging urgency and voltage sensitivity determine which EVs receive priority.

\begin{figure}[!t]
\centering
\resizebox{\columnwidth}{!}{%
\begin{tikzpicture}[
  box/.style={
    rectangle, rounded corners=2pt,
    draw, fill=white, align=center,
    text width=4.5cm,
    minimum height=0.75cm,
    font=\footnotesize, inner sep=4pt
  },
  dec/.style={
    diamond, draw, fill=white,
    aspect=2.3, align=center,
    font=\footnotesize, inner sep=2pt
  },
  arr/.style={
    -{Stealth[length=1.6mm]}, semithick
  },
  lab/.style={
    font=\scriptsize, fill=white, inner sep=1pt
  },
  phase/.style={
    font=\footnotesize\bfseries,
    anchor=north east,
    text width=4.3cm, align=right
  },
  conn/.style={
    circle, draw, fill=white,
    minimum size=5mm, inner sep=0pt,
    font=\scriptsize\bfseries
  }
]

\fill[orange!12]
  (-3.55,1.05) rectangle (8,-2.05);
\fill[cyan!10]
  (-3.55,-2.05) rectangle (8,-6.95);
\fill[green!10]
  (-3.55,-6.95) rectangle (8,-16.75);

\node[phase] at (7.75,0.75) {
  Phase 1: Normal operation
};
\node[phase] at (7.75,-2.35) {
  Phase 2:  Suspension\\
  and restoration
};
\node[phase] at (7.75,-7.25) {
  Phase 3: Backoff,\\
  hold and adapt
};

\node[box] (normal) at (0,0) {
  Admit all charging requests;\\
  measure critical-bus voltages
};

\node[dec] (vnormal) at (0,-1.3) {
  Undervoltage?
};

\node[conn] (ain) at (4,0) {A};

\node[box] (stop) at (0,-3.3) {
  Suspend controlled charging;\\
  set $N_{\mathrm{allow}}=0$;\\
  record $N_{\mathrm{viol}}$; set $\Delta N$
};

\node[box] (restore) at (0,-4.9) {
  Wait 15 min; update EV data;\\
  increase $N_{\mathrm{allow}}$ by $\Delta N$;\\
  dispatch EVs; measure voltages
};

\node[conn] (bin) at (-3.15,-4.9) {B};

\node[dec] (vrestore) at (0,-6.3) {
  Undervoltage?
};

\node[dec] (waiting) at (5,-4.9) {
  EVs\\waiting?
};

\node[conn] (aout) at (7.35,-4.9) {A};

\node[box] (backoff) at (0,-8.3) {
  Back off one charging group:\\
  $N_{\mathrm{allow}}\leftarrow
  \max\{0,N_{\mathrm{allow}}-\Delta N\}$;\\
  dispatch EVs; measure voltages
};

\node[dec] (vpersist) at (0,-9.8) {
  Undervoltage\\persists?
};

\node[conn] (dout) at (3.6,-9.8) {D};
\node[conn] (din)  at (3.4,-3.3) {D};

\node[box] (hold) at (0,-11.5) {
  Wait 15 min; update EV data;\\
  hold $N_{\mathrm{allow}}$;\\
  dispatch EVs; measure voltages
};

\node[conn] (cin) at (-3.15,-11.5) {C};

\node[dec] (vhold) at (0,-13) {
  Undervoltage?
};

\node[dec] (pending) at (0,-14.4) {
  EVs\\waiting?
};

\node[dec, aspect=1.8] (margin) at (5,-14.4) {
  Sufficient margin?\\
  $V_{\min}\ge V_{\lim}+\varepsilon$
};

\node[conn] (cout) at (7.6,-12.65) {C};
\node[conn] (bout) at (5,-16.2) {B};

\node[box, text width=3cm] (ret) at (0,-16) {
  Next control interval:\\
  return to normal operation
};

\draw[arr] (normal) -- (vnormal);
\draw[arr] (ain) -- (normal.east);

\draw[arr]
  (vnormal.west)
  -- node[lab,above] {No} (-2.85,-1.3)
  |- (normal.west);

\draw[arr]
  (vnormal)
  -- node[lab,right] {Yes}
  (stop);

\draw[arr] (stop) -- (restore);
\draw[arr] (bin) -- (restore.west);
\draw[arr] (restore) -- (vrestore);

\draw[arr]
  (vrestore.east)
  -| node[lab,pos=0.25,below] {No}
  (waiting.south);

\draw[arr]
  (waiting.west)
  -- node[lab,above] {Yes}
  (restore.east);

\draw[arr]
  (waiting.east)
  -- node[lab,above] {No}
  (aout);

\draw[arr]
  (vrestore)
  -- node[lab,right] {Yes}
  (backoff);

\draw[arr] (backoff) -- (vpersist);

\draw[arr]
  (vpersist)
  -- node[lab,right] {No}
  (hold);

\draw[arr]
  (vpersist.east)
  -- node[lab,above] {Yes}
  (dout);

\draw[arr] (din) -- (stop.east);

\draw[arr] (cin) -- (hold.west);
\draw[arr] (hold) -- (vhold);

\draw[arr]
  (vhold.west)
  -- node[lab,above] {Yes} (-2.85,-13)
  |- (backoff.west);

\draw[arr]
  (vhold)
  -- node[lab,right] {No}
  (pending);

\draw[arr]
  (pending.east)
  -- node[lab,above] {Yes}
  (margin.west);

\draw[arr]
  (pending)
  -- node[lab,right] {No}
  (ret);

\draw[arr]
  (margin.east)
  -| node[lab,pos=0.25,below] {No}
  (cout.south);

\draw[arr]
  (margin)
  -- node[lab,right] {Yes}
  (bout);

\end{tikzpicture}%
}
\caption{Flowchart of the VFBR controller.
Matching lettered connectors indicate continuation.
EV data updates include charging requests and priority rankings.
Dispatch selects the highest-ranked requesting EVs,
subject to the allowable number and current request count.
Voltages are measured after commands take effect.
Normal-operation monitoring occurs every 15 min;
connector A resumes normal operation at the next interval;
connector D restarts suspension and restoration when undervoltage persists after a backoff.}
\label{fig:flowchart}
\end{figure}
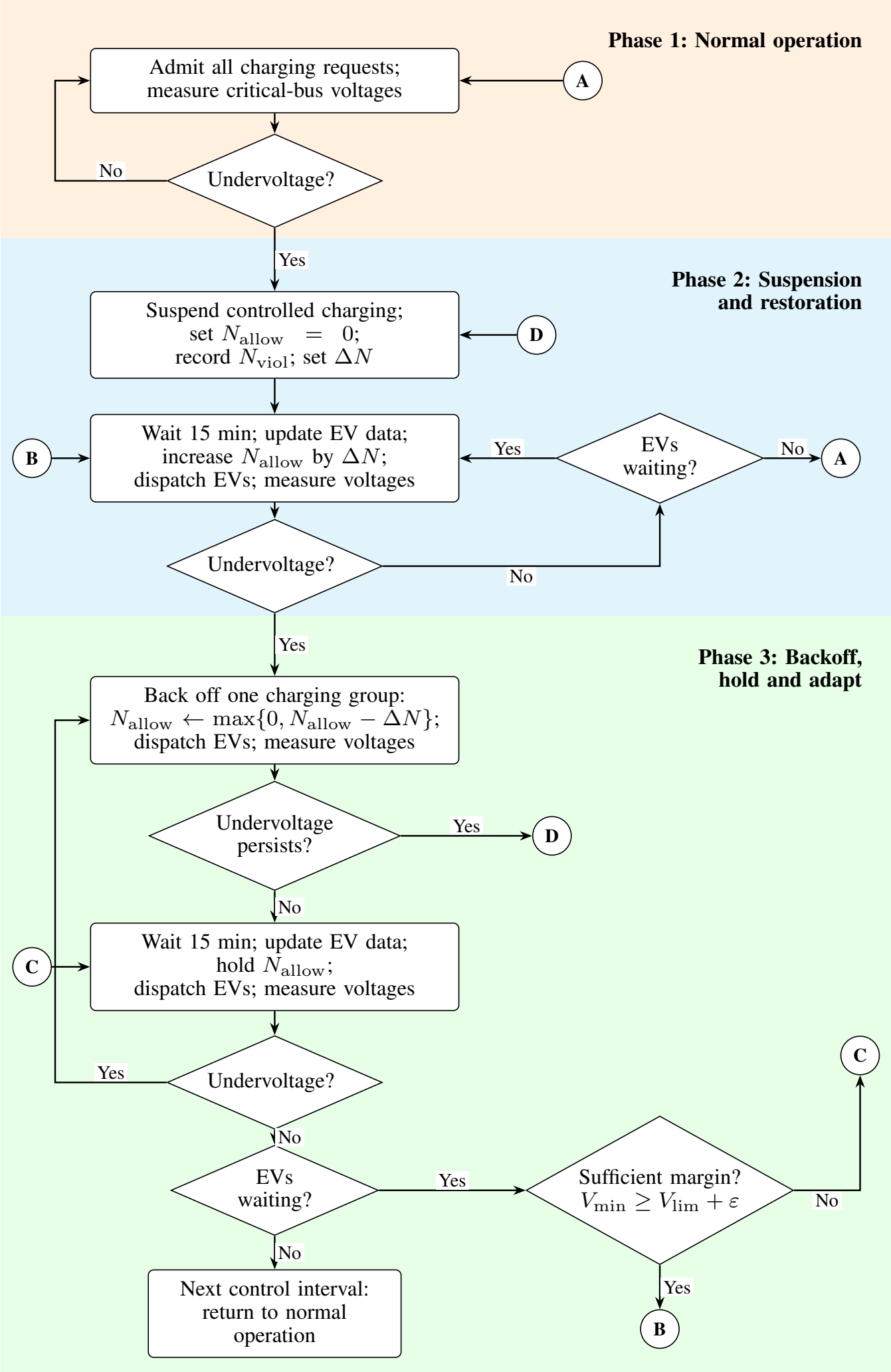

As illustrated in Fig.~\ref{fig:flowchart}, the framework operates in three phases. During normal operation, all charging requests are admitted while critical-bus voltages are monitored. When undervoltage is detected, the controller suspends all controlled EV charging at the next control interval and then progressively restores charging. During restoration, $N_{\mathrm{allow}}(t)$ is increased in increments of $\Delta N$, with the highest-priority EVs selected to charge. If undervoltage recurs, the allowable number is reduced by $\Delta N$. During the subsequent hold-and-adapt phase, the controller maintains the allowable number while updating EV requests, priorities, and charging assignments. Further undervoltage triggers additional curtailment, whereas sufficient voltage margin permits renewed restoration when EVs are waiting to charge.

The charging command for EV $i$ at interval $t$ is represented by $x_i(t)\in\{0,1\}$, where $x_i(t)=1$ enables charging and $x_i(t)=0$ suspends it. Its SOC evolves as
\begin{equation}
  SOC_i(t+1)=SOC_i(t)+
  \frac{\eta_i P_i^{\mathrm{ch}}\Delta t}{E_i^{\mathrm{bat}}}x_i(t),
  \label{eq:soc}
\end{equation}
where $\eta_i$ is the charging efficiency, $P_i^{\mathrm{ch}}$ is the rated charging power, and $E_i^{\mathrm{bat}}$ is the battery capacity. SOC is expressed as a fraction, and $\Delta t=15$~min when charging power and battery capacity are expressed in kW and kWh, respectively. EVs leave the charging-request set upon reaching their target SOC or departing.

\subsection{Voltage-Feedback Control of Concurrent Charging}

The VFBR controller adjusts the allowable number of concurrently charging EVs using critical-bus voltage measurements. Let $N_{\mathrm{req}}(t)$ denote the number of connected EVs requesting charging at interval $t$. As shown in Fig.~\ref{fig:flowchart}, the controller operates in three phases.

\textit{1) Normal Operation:} All charging requests are admitted, with $N_{\mathrm{allow}}(t)=N_{\mathrm{req}}(t)$. Critical-bus voltages and charging requests are updated every 15 min.

\textit{2) Suspension and Restoration:} Upon detecting undervoltage, the controller records the actual number of charging EVs, $N_{\mathrm{viol}}$, and suspends controlled charging by setting $N_{\mathrm{allow}}=0$. The restoration increment is $\Delta N=\left\lceil\alpha N_{\mathrm{viol}}\right\rceil$, where $\alpha\in(0,1]$ determines the group size. A smaller $\alpha$ provides finer adjustment but slower restoration, whereas a larger $\alpha$ restores more charging demand per step but may produce a larger voltage-limit overshoot. The increment remains fixed during the restoration and subsequent backoff process and is updated only at each new suspension.

Provided that voltage recovers after suspension, charging is restored in groups at 15-minute intervals:
\begin{equation}
  N_{\mathrm{allow}}(t+1)=
  \min\left\{N_{\mathrm{req}}(t+1),
  N_{\mathrm{allow}}(t)+\Delta N\right\}.
  \label{eq:restore}
\end{equation}
At each restoration step, charging requests and priorities are updated, the highest-ranked EVs are admitted, and voltages are measured after the commands take effect. Restoration continues until undervoltage recurs or all current requests are admitted. In the latter case, normal operation resumes at the next control interval.

\textit{3) Backoff, Hold, and Adapt:} If undervoltage recurs, the controller backs off one charging group:
\begin{equation}
  N_{\mathrm{allow}}\leftarrow
  \min\left\{N_{\mathrm{req}},
  \max\left[0,N_{\mathrm{allow}}-\Delta N\right]\right\}.
  \label{eq:stepback}
\end{equation}
Charging assignments are updated according to the priority ranking, and voltages are checked again. If undervoltage persists after a backoff, charging is suspended again, a new $N_{\mathrm{viol}}$ is recorded, and suspension and restoration restart.

Once voltage recovers, the controller holds the allowable number, capped by the current request count, while updating priorities and charging assignments every 15 min. Voltage monitoring continues because changes in residential demand or the locations of selected EVs can affect voltage even when the charging count remains unchanged. Any renewed undervoltage triggers another one-group backoff. If EVs are waiting to charge, restoration resumes only when
\begin{equation}
  V_{\min}(t)\ge V_{\mathrm{lim}}+\varepsilon,
  \label{eq:margin}
\end{equation}
where $\varepsilon>0$ introduces a voltage margin to discourage repeated switching between adjacent charging levels. Each restored group is followed by another voltage check. If all current requests are admitted without undervoltage, the controller returns to normal operation.

VFBR therefore adapts concurrent charging through voltage feedback without online optimization or forecasts of residential demand and future EV arrivals. It seeks an acceptable operating level rather than a guaranteed maximum feasible charging count. If undervoltage persists with all controlled charging suspended, charging remains disabled and additional voltage-control measures are required.

\subsection{EV Priority Formulation}

When charging requests exceed $N_{\mathrm{allow}}(t)$, requesting EVs are ranked by charging urgency and voltage sensitivity. Only connected EVs that have not reached their target SOC and have not departed are considered.

\textit{1) Charging Urgency:} Urgency represents the fraction of the remaining connection time required to complete charging:
\begin{equation}
  U_i(t)=\frac{E_i^{\mathrm{rem}}(t)}
  {\eta_i P_i^{\mathrm{ch}}T_i^{\mathrm{rem}}(t)},
  \label{eq:urgency}
\end{equation}
where $E_i^{\mathrm{rem}}(t)$ is the remaining battery energy needed to reach the target SOC, and $T_i^{\mathrm{rem}}(t)>0$ is the time until departure. Time is expressed in hours when energy and power are in kWh and kW, respectively. Higher $U_i(t)$ indicates greater urgency. A value of one requires uninterrupted charging until departure, while a value above one indicates that the target cannot be reached within the remaining time at the rated charging power.

\textit{2) Voltage Sensitivity:} Voltage sensitivity quantifies the voltage reduction at bus $m$ caused by incremental charging demand at bus $n$~\cite{wang2022}. Using perturb-and-observe power-flow analysis, it is estimated as
\begin{equation}
  S_{m,n}=
  \frac{V_m^{\mathrm{before}}-V_m^{\mathrm{after}}}
  {\Delta P_n},
  \label{eq:sensitivity}
\end{equation}
where $\Delta P_n>0$ is the added test load, and the numerator is the resulting voltage reduction at bus $m$. A larger $S_{m,n}$ indicates a greater adverse voltage impact.

The sensitivity matrix is precomputed offline using the feeder model at a representative operating point and stored for online allocation. At each control interval, the controller identifies the monitored critical bus with the lowest measured voltage, $m(t)=\arg\min_{m\in\mathcal{N}_{\mathrm{crit}}}V_m(t)$, and retrieves the corresponding matrix row. EV $i$, connected at bus $n(i)$, is assigned sensitivity $S_{m(t),n(i)}$. Thus, online prioritization requires only a lookup of precomputed sensitivities rather than repeated power-flow calculations. These sensitivities guide EV ranking, while measured voltage feedback determines whether charging should be restored or curtailed.

\textit{3) Combined Priority:} The priority score is
\begin{equation}
  \Pi_i(t)=w_uU_i(t)-w_s\widehat{S}_i(t),
  \qquad w_u,w_s\ge0,\quad w_u+w_s=1,
  \label{eq:priority}
\end{equation}
where $\widehat{S}_i(t)=S_{m(t),n(i)}/
\max_{j\in\mathcal{E}_{\mathrm{con}}(t)}S_{m(t),n(j)}$ is the normalized sensitivity, and $\mathcal{E}_{\mathrm{con}}(t)$ denotes connected EVs. The maximum sensitivity is assumed positive. Higher urgency raises priority, whereas greater voltage impact lowers it. Since urgency can exceed one, equal weights do not necessarily imply equal contributions from the two criteria.

The highest-ranked requesting EVs are admitted up to $N_{\mathrm{allow}}(t)$. Requests and priorities are updated at subsequent control intervals.

\section{Simulation Results}
\label{sec:case study}

\color{purple}
The proposed VFBR framework is evaluated on the unbalanced IEEE 123-bus distribution feeder modeled in OpenDSS over one week at a 15-minute resolution. In practical deployment, VFBR uses monitored critical-bus voltages as feedback. Here, OpenDSS power-flow solutions provide simulated voltage measurements by representing the feeder response to residential demand and EV charging commands. Power-flow calculations are part of the simulation environment, rather than an online computation required by the controller.

{\color{blue}Residential load profiles are obtained from actual smart-meter data,
while EV arrival times, departure times, and daily driving distances are modeled using the 2022 National Household Travel Survey (NHTS)~\cite{nhts2022}.} A total of 500 EVs are randomly assigned to residential loads on phase~A. Each EV has an 82-kWh battery and an identical 7.7-kW (240~V, 32~A) Level-2 charger. Charging losses are neglected ($\eta_i=1$), and the target SOC is 100\%.

The lower voltage limit is $V_{\mathrm{lim}}=0.95$~p.u., the restoration ratio is $\alpha=0.1$, and the voltage margin is $\varepsilon=0.005$~p.u. The voltage-sensitivity matrix is precomputed offline in OpenDSS using~\eqref{eq:sensitivity} and stored for priority allocation; online allocation only retrieves the corresponding sensitivity values.

\color{black}

Four allocation strategies are evaluated under the same VFBR control rules: \textit{random}, which ranks requesting EVs randomly; \textit{urgency-based}, which ranks EVs by decreasing $U_i(t)$ with $(w_u,w_s)=(1,0)$; \textit{sensitivity-based}, which prioritizes lower voltage sensitivities with $(w_u,w_s)=(0,1)$; and \textit{combined}, which ranks EVs by decreasing $\Pi_i(t)$ with $w_u\in\{0.25,0.5,0.75\}$ and $w_s=1-w_u$. Each strategy admits the highest-ranked requesting EVs up to $N_{\mathrm{allow}}(t)$. An uncontrolled-charging baseline is also included, in which each EV charges at its rated power from arrival until reaching its target SOC or departing, whichever occurs first.

\subsection{Comparison of EV Charging Request Satisfaction}

\begin{figure}[!t]
    \centering
    \includegraphics[width=0.9\columnwidth]{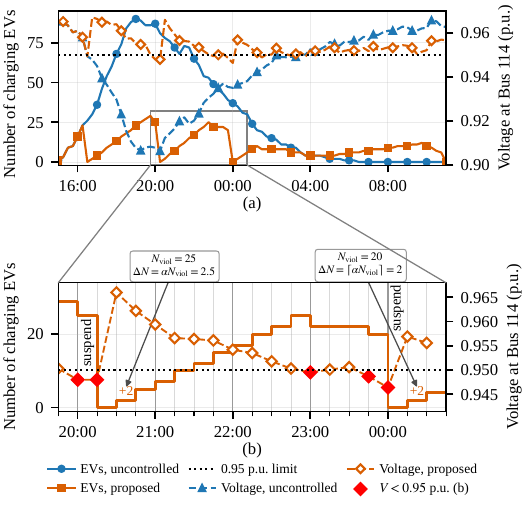}
    \caption{(a) Charging EVs and voltage at Bus~114 with uncontrolled charging and VFBR. (b) Enlarged boxed interval; red markers: $V<0.95$~p.u.}
    \label{fig:voltage_response}
\end{figure}

Uncontrolled charging meets all charging requirements but causes sustained undervoltage. In Fig.~\ref{fig:voltage_response}, approximately 90 EVs charge concurrently near 19:00, lowering the voltage at Bus~114 to 0.907~p.u. VFBR curtails peak charging and shifts demand overnight, maintaining voltage near or above 0.95~p.u. for most of the illustrated period.

\begin{figure}[!t]
    \centering
    \includegraphics[width=0.9\columnwidth]{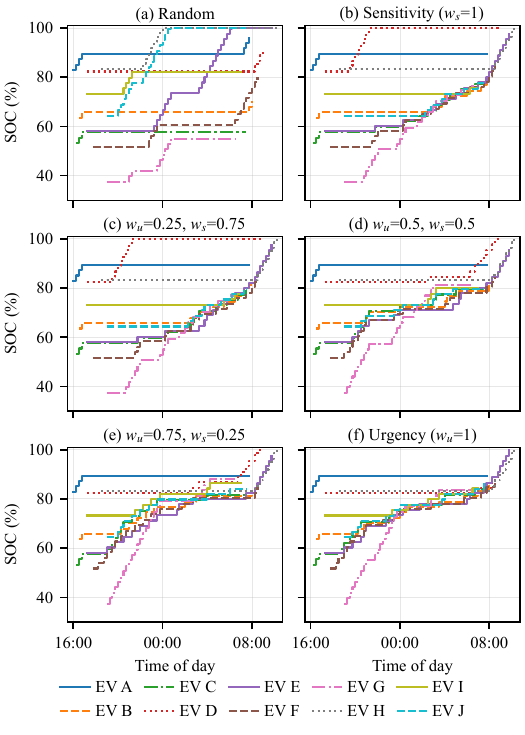}
    \caption{SOC trajectories of ten selected EVs (EV~A--EV~J) during one
    evening under the six charging strategies.}
    \label{fig:soc_trajectories}
\end{figure}

\begin{figure}[!t]
    \centering
    \includegraphics[width=0.9\columnwidth]{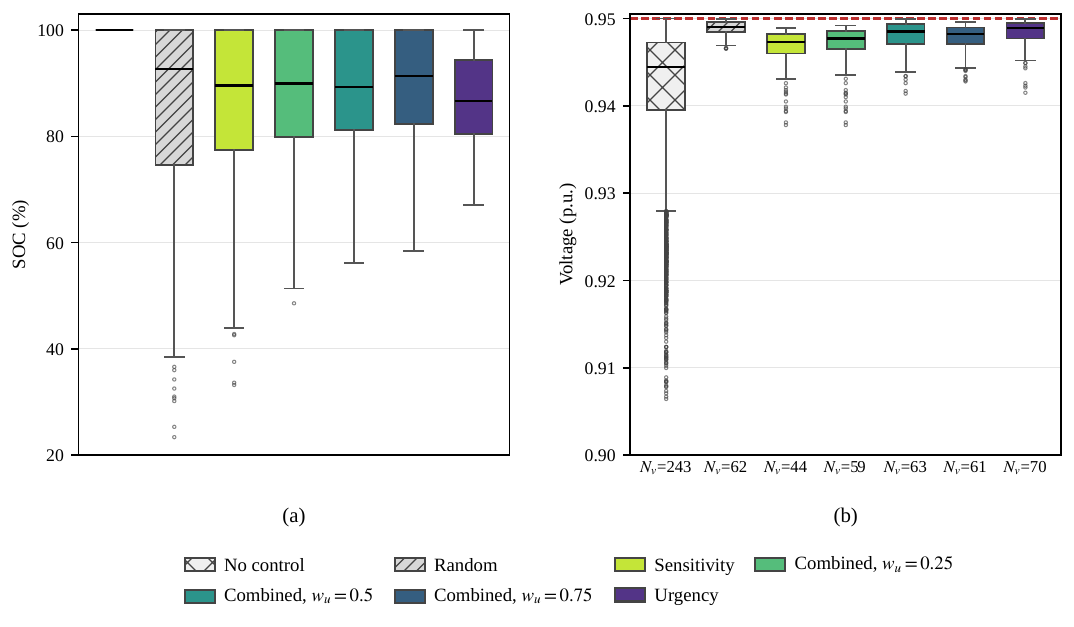}
    \caption{Distributions over the seven-day study period of (a) departure SOC and (b) bus voltages below 0.95~p.u. $N_{\mathrm{v}}$ denotes the number of time steps with at least one bus voltage below 0.95~p.u.}
    \label{fig:departure_soc}
\end{figure}

Figure~\ref{fig:soc_trajectories} illustrates the effect of allocation priorities. Random allocation leaves the initially most urgent EV~G at approximately 55\% SOC upon departure. Sensitivity-based allocation favors the less voltage-sensitive EV~D despite its low urgency, while increasing the urgency weight improves EV~G's departure SOC. Figure~\ref{fig:departure_soc} shows the corresponding aggregate tradeoff: urgency-based allocation achieves the highest daily minimum SOC but generally lower medians, whereas combined allocation improves minimum SOC while retaining higher medians on most days.

The two criteria serve complementary purposes: urgency prioritizes time-critical requests, while sensitivity favors locations that preserve voltage margin for other charging requests. Compared with urgency-only allocation, equal-weight allocation increases delivered energy by 12.6\%, raises mean departure SOC from 86.3\% to 88.5\%, improves Jain's index from 0.68 to 0.73, and reduces unserved sessions from 129 to 99. The minimum SOC, however, decreases from 67.1\% to 56.1\%. Increasing $w_u$ to 0.75 achieves the highest delivered energy and mean SOC among the evaluated strategies, illustrating the benefit of balancing charging urgency and voltage impact.

\subsection{Overall Performance Comparison}

As shown in Fig.~\ref{fig:voltage_response}, VFBR first suspends controlled EV charging to rapidly relieve undervoltage, then restores charging group by group using measured voltage feedback. Priority rankings determine which EVs are admitted, allowing charging service to recover within the available voltage margin without online optimization.

\begin{figure}[!h]
\centering
\includegraphics[width=0.84\columnwidth]{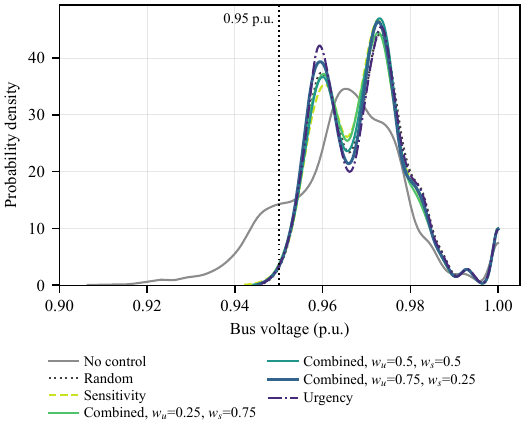}
\caption{Kernel density estimate of the voltages at all buses during the
charging period (16:00--08:00).}
\label{fig:vmin_kde}
\end{figure}

Figure~\ref{fig:vmin_kde} compares bus-voltage distributions across strategies. During the charging period, VFBR reduces the proportion of bus-time samples below 0.95~p.u. from 15\% to less than 1\% and raises the minimum voltage from 0.906 to above 0.938~p.u. No overvoltage occurs. All controlled strategies achieve similar voltage performance, while their allocation priorities produce different charging outcomes.

Table~\ref{tab:comparison} summarizes charging performance, with fairness measured by Jain's index~\cite{jain1984}. Random allocation yields the lowest delivered energy, minimum departure SOC, and fairness. Urgency-based allocation achieves the highest minimum SOC (67.1\%) but delivers 10.13~MWh and leaves 129 sessions unserved. Combined allocation with $(w_u,w_s)=(0.75,0.25)$ achieves the highest delivered energy (11.90~MWh), mean departure SOC (89.4\%), and fairness (0.75), with the fewest unserved sessions (98). Thus, balancing urgency and voltage sensitivity improves aggregate charging service, whereas urgency-only allocation better protects the minimum departure SOC.

\begin{table}[!t]
\centering
\setlength{\arrayrulewidth}{1pt}
\caption{Comparison of Allocation Strategies}
\label{tab:comparison}
\footnotesize
\setlength{\tabcolsep}{4pt}
\begin{tabular}{lccccc}
\hline
\textbf{Strategy} & \textbf{Energy} & \textbf{Mean SOC} & \textbf{Min SOC} & \textbf{Not} & \textbf{Jain} \\
 & (MWh) & (\%) & (\%) & \textbf{served} & \\
\hline
Random       & 9.52  & 85.2 & 23.4 & 168 & 0.64 \\
Sensitivity  & 10.87 & 87.6 & 33.2 & 121 & 0.68 \\
$w_u$=0.25   & 11.34 & 88.4 & 48.6 & 114 & 0.71 \\
$w_u$=0.5    & 11.41 & 88.5 & 56.1 & 99  & 0.73 \\
\rowcolor[gray]{0.92}
\textbf{$w_u$=0.75 }  & \textbf{11.90} & \textbf{89.4} & 58.4 & \textbf{98} & \textbf{0.75} \\
Urgency      & 10.13 & 86.3 & \textbf{67.1} & 129 & 0.68 \\
\hline
\end{tabular}
\end{table}

\begin{table}[!t]
\centering
\setlength{\arrayrulewidth}{1pt}
\caption{Effect of the Restoration Ratio ($w_u=0.75$, One Week)}
\label{tab:alpha}
\footnotesize
\setlength{\tabcolsep}{4pt}
\begin{tabular}{cccccc}
\hline
$\boldsymbol{\alpha}$ & \textbf{Energy} & \textbf{Mean SOC} & \textbf{Min SOC} & \textbf{Not} & \textbf{Jain} \\
 & (MWh) & (\%) & (\%) & \textbf{served} & \\
\hline
0.1 & 11.90 & 89.4 & 58.4 & 98 & 0.75 \\
\rowcolor[gray]{0.92}
\textbf{0.3} & \textbf{12.72} & \textbf{90.9} & \textbf{66.0} & \textbf{66} & \textbf{0.79} \\
0.5 & 11.94 & 89.5 & 56.2 & 77 & 0.77 \\
\hline
\end{tabular}
\end{table}
\begin{figure}[!t]
\centering
\includegraphics[width=0.84\columnwidth]{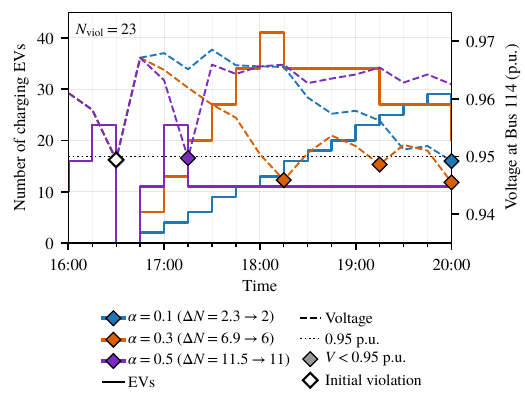}
\caption{Charging EVs (solid) and Bus 114 voltage (dashed) after the first suspension for different $\alpha$ ($w_u=0.75$). Filled markers show voltages below 0.95~p.u. for each $\alpha$; the open marker is the initial violation.}
\label{fig:alpha_ev}
\end{figure}

{\color{blue}The strategies in Table~\ref{tab:comparison} are compared at a fixed $\alpha=0.1$; the best-performing one, $(w_u,w_s)=(0.75,0.25)$, is then simulated with $\alpha=0.3$ and 0.5 over one week (Table~\ref{tab:alpha}). Increasing $\alpha$ to 0.3 raises the delivered energy from 11.90 to 12.72~MWh and reduces the unserved sessions from 98 to 66. As shown in Fig.~\ref{fig:alpha_ev}, a larger $\alpha$ restores more EVs per step and brings the first post-restoration violation earlier. With $\alpha=0.5$, a large backoff holds few EVs for several hours, reducing the energy to 11.94~MWh. Thus, a moderate $\alpha$ balances restoration speed and backoff size.}

\color{purple}
\section{Conclusion}
\label{sec:Conclusion}

This paper proposed VFBR to mitigate EV-induced undervoltage through charging suspension and group-wise restoration using voltage feedback, without demand forecasts or online optimization. Charging priorities combine urgency and precomputed voltage sensitivity to balance user needs against feeder voltage constraints. One-week simulations on the unbalanced IEEE 123-bus feeder reduced charging-period undervoltage samples from 15\% to less than 1\% and raised the minimum voltage from 0.906 to above 0.938~p.u. Urgency-only allocation achieved the highest minimum departure SOC, while combined allocation with $(w_u,w_s)=(0.75,0.25)$ delivered 25\% more energy than random allocation and achieved the highest mean departure SOC and fairness. {\color{blue}Among the tested restoration ratios, $\alpha=0.3$ delivered the most energy, whereas $\alpha=0.5$ caused large backoffs that reduced charging service}. Future work will assess different EV penetration levels, feeder conditions, control intervals, and restoration step sizes.
\color{black}

\vspace{12pt}

\begin{thebibliography}{00}

\bibitem{b2}
K. Clement-Nyns, E. Haesen, and J. Driesen,
``The impact of charging plug-in hybrid electric vehicles on a residential
distribution grid,'' \emph{IEEE Trans. Power Syst.}, vol. 25, no. 1,
pp. 371--380, Feb. 2010.

\bibitem{pohl2025}
E. Pohl, S. Ghosh, P. Paudyal, M. Zhang, and M. J. Kisacikoglu,
``Evaluating load coincidence and distribution grid impacts of residential
EV charging,'' in \emph{Proc. IEEE Power \& Energy Society General Meeting
(PESGM)}, 2025, pp. 1--5.

\bibitem{banol2023}
N. Ba\~{n}ol Arias, C. Sabill\'{o}n, J. F. Franco, J. Quir\'{o}s-Tort\'{o}s,
and M. J. Rider, ``Hierarchical optimization for user-satisfaction-driven
electric vehicles charging coordination in integrated MV/LV networks,''
\emph{IEEE Syst. J.}, vol. 17, no. 1, pp. 1247--1258, Mar. 2023.

\bibitem{ucer2021}
E. Ucer, M. C. Kisacikoglu, and M. Yuksel,
``Decentralized additive increase and multiplicative decrease-based electric
vehicle charging,'' \emph{IEEE Syst. J.}, vol. 15, no. 3, pp. 4272--4280, Sep. 2021.

\bibitem{quiros2016}
J. Quir\'{o}s-Tort\'{o}s, L. F. Ochoa, S. W. Alnaser, and T. Butler,
``Control of EV charging points for thermal and voltage management of LV
networks,'' \emph{IEEE Trans. Power Syst.}, vol. 31, no. 4,
pp. 3028--3039, Jul. 2016.

\bibitem{dreucci2025}
D. Dreucci, Y. Yu, G. R. Chandra Mouli, A. Shekhar, and P. Bauer,
``Centralised distribution grid congestion management through EV charging
control considering fairness and priority,'' \emph{Appl. Energy},
vol. 384, Art. no. 125417, 2025.

\bibitem{wang2022}
L. Wang, A. Dubey, A. H. Gebremedhin, A. K. Srivastava, and N. Schulz,
``MPC-based decentralized voltage control in power distribution systems
with EV and PV coordination,'' \emph{IEEE Trans. Smart Grid},
vol. 13, no. 4, pp. 2908--2919, Jul. 2022.

\bibitem{nhts2022}
Federal Highway Administration, ``2022 National Household Travel Survey,'' U.S. Department of Transportation, Washington, DC, 2022. [Online]. Available: https://nhts.ornl.gov

\bibitem{jain1984}
R. Jain, D.-M. Chiu, and W. Hawe, ``A quantitative measure of fairness
and discrimination for resource allocation in shared computer systems,''
DEC Research Report TR-301, 1984.

\end{thebibliography}
\end{document}